\documentclass[pra,aps,twocolumn,superscriptaddress,nofootinbib,a4]{revtex4}
\usepackage{amsmath,epsfig}
\usepackage{graphicx}
\usepackage{graphics}

\begin{document}

\title{Entanglement Sharing and Decoherence in the Spin-Bath}
\author{Christopher M. Dawson}
\email{dawson@physics.uq.edu.au}
\affiliation{School of Physical Sciences, The University of Queensland, St Lucia, QLD 4072, Australia}
\author{Andrew P. Hines}
\email{hines@physics.uq.edu.au}
\affiliation{Centre for Quantum Computer Technology, The University of Queensland, St Lucia, QLD 4072, Australia}
\affiliation{School of Physical Sciences, The University of Queensland, St Lucia, QLD 4072, Australia}
\author{Ross H. McKenzie}
\affiliation{School of Physical Sciences, The University of Queensland, St Lucia, QLD 4072, Australia}
\author{G.J. Milburn}
\affiliation{Centre for Quantum Computer Technology, The University of Queensland, St Lucia, QLD 4072, Australia}
\affiliation{School of Physical Sciences, The University of Queensland, St Lucia, QLD 4072, Australia}

\begin{abstract}
The monogamous nature of entanglement has been illustrated by the
derivation of entanglement sharing inequalities - bounds on the amount of
entanglement that can be shared amongst the various parts of a
multipartite system. Motivated by recent studies of decoherence, we
demonstrate an interesting manifestation of this phenomena that arises in
system-environment models where there exists interactions between the
modes or subsystems of the environment. We investigate this phenomena in
the spin-bath environment, constructing an entanglement sharing inequality
bounding the entanglement between a central spin and the environment in
terms of the pairwise entanglement between individual bath spins. The
relation of this result to decoherence will be illustrated using simplified system-bath
models of decoherence.
\end{abstract}
\maketitle

While entanglement is argued to be the distinguishing feature of quantum
computers, responsible for their power \cite{Bennett00a}, it is also the
source of one of the major obstacles in their construction. {\it
Decoherence}, the process by which a quantum superposition state decays
into a classical, statistical mixture of states, is caused by entangling
interactions between the system and its environment \cite{Zurek91a}.
Somewhat paradoxically, the quantum entanglement between a system and its
environment induces classicality in the system. While it is still a
contentious topic as to whether quantum computation will be possible in
the face of decoherence, Zurek \cite{Zurek82a} has demonstrated that
decoherence is necessary to facilitate the measurement of a quantum
system. Understanding decoherence lies at the heart of measurement,
quantum information processing and, more fundamentally, the transition
from the quantum to the classical world.

The road to studying decoherence by explicitly modeling
system-environment interactions has led to simple models of the quantum
environment. Environments can be modeled as either baths of harmonic
oscillators \cite{Weiss99a} or spins (with spin-$\frac{1}{2}$) argued to
represent distinct types of environmental modes \cite{Prokofev00a}. The
simplest system-environment models consist of a central spin (or qubit)
coupled to the environment - i.e. the spin-boson model \cite{Weiss99a} -
which has applications to the decoherence of qubits for quantum
information processing.


Decoherence of a spin-$\frac{1}{2}$ particle at low temperatures may be
conveniently modeled by the `central spin' model , which couples a
central spin-$\frac{1}{2}$ particle $\mathcal{S}$ to a spin-bath
$\mathcal{B}$ of $N$ spin-$\frac{1}{2}$ particles. A typical Hamiltonian
for this model may be written in the form
\begin{equation}
    \label{eqn:decoherence-ham}
    H = H_\mathcal{S} + H_\mathcal{B} + H_\mathcal{SB},
\end{equation}
where $H_\mathcal{S}$, $H_\mathcal{B}$ are the internal Hamiltonians of
the central spin and spin-bath respectively, and $H_\mathcal{SB}$ is the
coupling term. Denote the state of the system-environment at time $t$ by
$\rho_\mathcal{SB}(t)$. Initially  at $t=0$ we take the central spin
$\mathcal{S}$ to be in a pure state, uncorrelated with the bath. That
is,
\begin{equation}
    \rho_\mathcal{SB}(0) = |\psi\rangle_\mathcal{S}\langle \psi | \otimes
    \rho_\mathcal{B}(0)
\end{equation}
for some initial state of the bath $\rho_\mathcal{B}(0)$. Typically
$\rho_\mathcal{B}(0)$ is taken to be a thermal state of the Hamiltonian
$H_\mathcal{B}$, or at low temperatures the ground state.

As the system evolves under $H$ the central spin becomes coupled to the
bath, and its reduced density matrix $\rho_\mathcal{S}(t)$ at later times
is no longer pure. The central spin is said to have decohered, and the
amount of decoherence is typically quantified by the von Neumann entropy
of its reduced density matrix $S(\rho_\mathcal{S}(t))$.

More recently interactions between modes within the bath itself have been considered
\cite{Tessieri02a,Paganelli02a,Lucamarini04a}, which allow for
appreciable correlations, such as entanglement, to arise between the modes
of the bath.

In \cite{Tessieri02a}, Tessieri and Wilkie introduced coupling terms
between spins in the bath Hamiltonian $H_\mathcal{B}$ and, taking the
initial state of the bath as a thermal state of $H_\mathcal{B}$, found that
this resulted in a suppression of the decoherence
$S(\rho_\mathcal{S}(t))$. The amount of suppression increased as the
effective energy scale of $H_\mathcal{B}$ increased relative to that of
$H_\mathcal{SB}$, ultimately to the point where decoherence was negligible
even after long times. This is somewhat surprising, as even small
couplings $H_\mathcal{B}$ would usually be expected to eventually result
in complete decoherence of the central spin. In this article we aim
to demonstrate that this suppression effect may be understood to be a
consequence of \emph{entanglement-sharing}, and that it will be common to
any central spin whose environment maintains appreciable internal
entanglement while involving in time.


A simple example of such a system is a single
spin in a bath of spins with antiferromagnetic interactions between
them. In the absence of the spin the ground state of the N bath spins
 would be something like a spin singlet which
is highly entangled. If the single spin interacts antiferromagnetically
with the bath spins all it can do is flip individual spins in the bath. The total spin has to be conserved and hence will have a  value of order $1/2$. If the bath is initialized in such a state, it will remain highly entangled throughout its interaction with system spin.

Entanglement sharing refers to a striking difference between classical and
quantum correlations --- the latter may not be shared arbitrarily amongst
several observables. The connection with decoherence is readily seen in a
system three spin-$\frac{1}{2}$ particles, labeled $\mathcal{S}, B_1, B_2$
respectively. It has been shown \cite{Coffman00a} that entanglement
between $B_1$ and $B_2$ limits the individual and collective entanglement
they may have with $\mathcal{S}$. If a state of the system $\rho(t)$ is evolving
under a Hamiltonian such as (\ref{eqn:decoherence-ham}), and moreover if the
`bath' $B_1 B_2$ maintains appreciable entanglement, then it follows there
is a restriction on the entanglement between the `central spin' $\mathcal{S}$ and
$B_1 B_2$. For pure states this equivalent to a restriction on the amount
that $\mathcal{S}$ may decohere. For mixed states we must also bound the classical
correlations between $\mathcal{S}$ and $B_1 B_2$ which may be done using a recent
result of Koashi \emph{et al}. \cite{Koashi03a}. Entanglement between $B_1$ and
$B_2$ thus suppresses all correlations between the central spin and
the bath.

The situation becomes far more complicated for spin-baths of $N$
particles. The main difficulty is the plethora of different types of
entanglement which exist in these baths, and the absence of good
entanglement measures for them. To overcome this difficulty we will assume
there is some symmetry in the Hamiltonians $H_{\mathcal{S}}$ and
$H_{\mathcal{SB}}$. If the initial bath state $\rho_\mathcal{B}(t)$ is
taken to be a thermal or eigenstate of $H_\mathcal{B}$ then the reduced
state of the bath $\rho_\mathcal{B}(t)$ at later times will also obey this
symmetry. For example, the simplest case is that considered by Tessieri
and Wilkie where $H_\mathcal{SB}$ and $H_\mathcal{B}$ are completely
symmetric. Here the pairwise entanglement between any two bath spins is
the same, allowing us to quantify the bath entanglement by a single
parameter.

In this paper we will obtain an entanglement-sharing inequality relating
the entanglement between a central spin and a completely symmetric
spin-bath to the pairwise entanglement in the bath. This inequality is
applicable to both pure and mixed states, and is sufficient to restrict
decoherence where $\rho_\mathcal{SB}(t)$ is pure. We will then illustrate
this damping effect in a simple model of decoherence originally proposed
by Zurek \cite{Zurek82a} and the Tessieri and Wilkie model \cite{Tessieri02a}. To conclude we will discuss possible extensions
of this result to the bounding of classical correlations between the central spin and the bath.

To begin, let $\mathcal{S}$ be a central spin-$\frac{1}{2}$ particle and
$\mathcal{B} = B_1 B_2 \ldots B_N$ a completely symmetric spin-bath. As indicated above, the symmetry
implies that the entanglement between any pair of bath spins $B_i, B_j$ is
the same, allowing us to use a single parameter as a measure of bath entanglement. This entanglement will be called the
\emph{intra-bath} entanglement, while the entanglement between the central
spin and the bath will be called the \emph{system-bath} entanglement. To quantify these we will make use of a
measure known as the \emph{tangle} \cite{Coffman00a} whose definition we
now briefly recall. For the reduced density matrix $\rho_{B_i B_j}$ of a
pair of bath spins $B_i,B_j$ define the spin-flipped density matrix
\begin{equation}
    \label{eqn:tilde}
    \tilde{\rho}_{B_i B_j} = (\sigma_y \otimes \sigma_y) \rho_{B_i B_j}^* (\sigma_y \otimes \sigma_y).
\end{equation}
The asterix denotes complex conjugation in the standard basis and
$\sigma_y$ is the Pauli $Y$ matrix. The matrix $\rho_{B_i B_j}
\tilde{\rho}_{B_i B_j}$ can be shown to have real non-negative
eigenvalues, and we write their square roots in decreasing order as
$\lambda_1, \lambda_2, \lambda_3, \lambda_4$. The tangle between $B_i$ and
$B_j$ is then defined as
\begin{equation}
    \label{eqn:tangle}
    \tau_{B_i|B_j} = \left( \max \left\{ 0,\lambda_1 - \lambda_2 - \lambda_3 - \lambda_4 \right\} \right)^2 .
\end{equation}
This expression is for two spin-$\frac{1}{2}$ particles, however the
tangle between the central spin $\mathcal{S}$ and the bath $\mathcal{B}$ is also
well-defined for pure states of the combined system. The key point is
that, because $\mathcal{S}$ is a spin-$\frac{1}{2}$ particle, only two dimensions of
the bath state-space are required to expand the pure state in its Schmidt
decomposition. The bath may therefore be imagined as a single
spin-$\frac{1}{2}$ particle, with the tangle defined as before.
Eq.~(\ref{eqn:tangle}) can be further simplified for pure states so that
the system-bath tangle is
\begin{equation}
    \label{eqn:puretangle}
    \tau_{\mathcal{S}|\mathcal{B}} = 4 \det{\rho_\mathcal{S}} .
\end{equation}
For further properties of the tangle, in particular its validity as an
entanglement measure, we refer the reader to
\cite{Coffman00a,Wootters98a}.

Since all the pairwise intra-bath tangles are the same we write
$\tau_B \equiv \tau_{B_i|B_j}$ for all $i,j$. Our aim is to show how this
$\tau_B$ constrains the system-bath tangle $\tau_{\mathcal{S}|\mathcal{B}}$. We will
first consider the simplest case of pure states for an $N=2$ bath, since
much is known about states of three spin-$\frac{1}{2}$ particles.
Intuition built in this case will enable us to derive a related inequality
for pure states of arbitrary sized baths.

For the two-spin bath, it was shown in \cite{Coffman00a} and \cite{Dur00a}
that there are two distinct types of entanglement between $\mathcal{S}$ and $B_1
B_2$. $\mathcal{S}$ can be entangled with the spins $B_1$ and $B_2$ individually, or
with the bath $B_1 B_2$ as a whole. The latter type is quantified by the
\emph{three-tangle} which we denote by $\tau_{\mathcal{S}|B_1|B_2}$. The total
entanglement between $\mathcal{S}$ and $\mathcal{B}$ can now be written as
\begin{equation}
    \label{eqn:system-bath-components}
    \tau_{\mathcal{S}|\mathcal{B}} = \tau_{\mathcal{S}|B_1} + \tau_{\mathcal{S}|B_2} + \tau_{\mathcal{S}|B_1|B_2} .
\end{equation}
The three-tangle is invariant under permutations of the three spins, and
may be written alternatively as
\begin{eqnarray}
    \label{eqn:3tangle1}
    \tau_{\mathcal{S}|B_1|B_2} & = & \tau_{\mathcal{S}|B_1 B_2} - \tau_{\mathcal{S}|B_1} - \tau_{\mathcal{S}|B_2} \\
    \label{eqn:3tangle1}
    \tau_{\mathcal{S}|B_1|B_2} & = &\tau_{B_1|\mathcal{S} B_2} - \tau_B - \tau_{B_1|\mathcal{S}}.
\end{eqnarray}
A simple consequence of this, together with the fact that the tangle is a
positive quantity less than or equal to one, is
\begin{equation}
    \tau_B + \tau_{\mathcal{S}|B_1|B_2} \leq 1 .
\end{equation}
This inequality says that the intra-bath entanglement plus the
three-tangle part of the system-bath entanglement is always less than $1$.
On the other hand, the sum of $\tau_{B} + \tau_{\mathcal{S}|B_1} + \tau_{\mathcal{S}|B_2}$ can
be greater than $1$ --- it can take any value up to and including $4/3$
\cite{Dur00a}. This suggests that intra-bath entanglement has a stronger
damping effect on the three-tangle component of $\tau_{\mathcal{S}|\mathcal{B}}$ than
it does on the pairwise tangle component. We will therefore assume that,
for a fixed intra-bath tangle, a maximum system-bath entanglement is
obtained when $\tau_{\mathcal{S}|B_1|B_2} = 0$. That is, when it is composed
entirely of the pairwise components in
Eq.~(\ref{eqn:system-bath-components}).

States of the $\mathcal{S}B_1 B_2$ system with $\tau_{\mathcal{S}|B_1|B_2} = 0$ are equivalent
under local unitary operations to so called $W$-class states of the form
\begin{eqnarray}
    \label{eqn:wclass}
    |\psi\rangle & = & a|{\uparrow}\rangle_\mathcal{S} |{\uparrow\downarrow}\rangle_B + b|{\uparrow}\rangle_\mathcal{S} |{\downarrow\uparrow}\rangle_B \\ \nonumber
& & + \,\, c|{\downarrow}\rangle_\mathcal{S} |{\uparrow\uparrow}\rangle_B +
d|{\uparrow}\rangle_\mathcal{S} |{\uparrow\uparrow} \rangle_B
\end{eqnarray}
where $a,b,c,d$ are real and non-negative \cite{Dur00a,Carteret00a} and
$a^2+b^2+c^2+d^2=1$. The tensor factors in each term refer to the state of
the central spin and of the two bath spins respectively. It is a simple
matter to calculate the relevant tangles from Eqs~(\ref{eqn:tangle},
\ref{eqn:puretangle})
\begin{eqnarray}
    \label{eqn:wtangles1}
    \tau_B & = & 4 a^2b^2  \\
    \label{eqn:wtangles2}
    \tau_{\mathcal{S}|\mathcal{B}} & = & 4 (a^2+b^2)c^2 .
\end{eqnarray}
We will solve the equivalent, and as it turns out slightly easier, problem
of maximizing $\tau_B$ for fixed $\tau_{\mathcal{S}|\mathcal{B}} = T$. That is, we
must maximize
\begin{equation}
    g(a,b,c,d) = 4 a^2 b^2
\end{equation}
subject to the constraints
\begin{eqnarray}
    \label{eqn:wcon1}
    F_1(a,b,c,d) & = & 4(a^2 + b^2) c^2 - T = 0 \\
    \label{eqn:wcon2}
    F_2(a,b,c,d) & = & a^2 + b^2 + c^2 + d^2 - 1 = 0 .
\end{eqnarray}

This can be solved by the method of Lagrange multipliers, and we find the
maximum $\tau_B$ is given by
\begin{equation}
    \label{eqn:2def}
    \tau_B = \frac{1}{4}\left( 1 + \sqrt{1-\tau_{\mathcal{S}|\mathcal{B}}} \right)^2 .
\end{equation}
The corresponding entanglement-sharing inequality for the system-bath and
intra-bath tangles is then
\begin{equation}
    \label{eqn:es-ineq2}
    \tau_{\mathcal{S}|\mathcal{B}} \leq \left\{
        \begin{array}{cc}
            1 & \,\,\,\,\,\,\, \tau_B \leq \frac{1}{4} \\
            4\left(\sqrt{\tau_B}-\tau_B\right) & \,\,\,\,\,\,\, \tau_B \geq \frac{1}{4}.
        \end{array} \right.
\end{equation}
For values of the intra-bath tangle less than $1/4$ the system and the
bath may be maximally entangled. As $\tau_B$ increases however, we find
that $\tau_{\mathcal{S}|\mathcal{B}}$ falls in an approximately linear fashion, and
is $0$ when the intra-bath tangle is at a maximum. This confirms our
expectation that strong quantum correlations in the environment limit
decoherence effects, at least for pure states of the combined system.


We saw above that the three-tangle component of the system-bath
entanglement was more strongly limited by the intra-bath entanglement than
the pairwise components $\tau_{\mathcal{S}|B_1},\tau_{\mathcal{S}|B_2}$. In the case of an
$N$-spin bath it seems reasonable that we should expect the same, this
time potentially for three-party and other higher order quantum
correlations between $\mathcal{S}$ and the bath. We will therefore assume that
analogues of the $W$-class states are able to achieve maximum system-bath
entanglement for a given intra-bath entanglement. An inequality similar to
Eq.~(\ref{eqn:es-ineq2}) follows from this assumption and has been
confirmed numerically for small values of $N$.

An analogue of a $W$-class state should ideally be one where the system is
only entangled with each of the bath spins individually. We will use a
generalization of the states $(\ref{eqn:wclass})$ given by
\begin{eqnarray}
    \label{eqn:wstateN}
    |W\rangle & = & a_1 |{\uparrow}\rangle_\mathcal{S} |{\uparrow \uparrow \cdots \uparrow \uparrow \downarrow} \rangle_B
        + \,\, a_2 |{\uparrow}\rangle_\mathcal{S} |{\uparrow \uparrow \cdots \uparrow \downarrow \uparrow} \rangle_B \nonumber \\
        & & + \,\, \cdots \nonumber \\
        & & + \,\, a_N |{\uparrow}\rangle_\mathcal{S} |{\downarrow \uparrow \cdots \uparrow \uparrow \uparrow} \rangle_B
        + \,\, c |{\downarrow}\rangle_\mathcal{S}  |{\uparrow \uparrow \cdots \uparrow \uparrow \uparrow} \rangle_B
        \nonumber \\
        & & + \,\, d  |{\uparrow}\rangle_\mathcal{S}  |{\uparrow \uparrow \cdots \uparrow \uparrow \uparrow} \rangle_B
\end{eqnarray}
for real $a_i,c,d$ where $\sum_{i=1}^N a_i^2 + c^2 + d^2 = 1$. $a_i$ is
the coefficient of the state where the $i$th bath spin is down. From
Eqs~(\ref{eqn:tangle},\ref{eqn:puretangle}) we find that the tangle
between any pair of bath-spins is given by
\begin{equation}
    \tau_{B_i|B_j} = 4 a_i^2 a_j^2 ,
\end{equation}
and the tangle between the central spin and the bath is given by
\begin{equation}
    \tau_{\mathcal{S}|\mathcal{B}} = 4 c^2 \sum_{i=1}^N a_i^2 .
\end{equation}
The symmetry constraint implies that $a_i = a_j = a$ for all $i,j \leq N$,
and it follows that
\begin{eqnarray}
    \tau_{B} & = & \tau_{B_i|B_j} = 4 a^4 \\
    \tau_{\mathcal{S}|\mathcal{B}} & = & 4N a^2 c^2 .
\end{eqnarray}
Fixing $\tau_{\mathcal{S}|\mathcal{B}} = D$ we can maximize $\tau_{B}$ as we did for
the $N=2$ case, and subsequently obtain a maximum $\tau_B$ at
\begin{equation}
    \label{eqn:es-preineqN}
    \tau_B = \frac{1}{N^2}\left( 1 + \sqrt{1-\tau_{\mathcal{S}|\mathcal{B}}} \right)^2
\end{equation}
with the corresponding entanglement-sharing inequality
\begin{equation}
    \label{eqn:es-ineqN}
    \tau_{\mathcal{S}|\mathcal{B}} \leq \left\{
        \begin{array}{cc}
            1 & \,\,\,\,\,\,\, \tau_B \leq \frac{1}{N^2} \\
            N\left(2\sqrt{\tau_B}-N\tau_B\right) & \,\,\,\,\,\,\, \tau_B \geq \frac{1}{N^2}.
        \end{array} \right.
\end{equation}

This inequality is identical to Eq.~(\ref{eqn:es-ineq2}) up to a
dimensional scaling. Note that the maximum possible pairwise tangle for a
symmetric bath of $N$ spin-$\frac{1}{2}$ particles has been shown to be
$4/N^2$ \cite{Koashi00a}, and that the system-bath tangle falls to $0$ for
this value of $\tau_B$.

Of course, we have only demonstrated this inequality for the $W$-class
states Eq.~(\ref{eqn:wstateN}). To verify the inequality numerically for
small values of $N$ we calculated $\tau_{\mathcal{S}|\mathcal{B}}$ and $\tau_{B}$ for
random states having the appropriate bath symmetry. A sample size of $1
\times 10^7$ was used, and to reduce the sample space we used the
generalized Schmidt decomposition \cite{Carteret00a}. No violations of
Eq.~(\ref{eqn:es-ineqN}) were found for $N \leq 5$.


The extension of Eq.~(\ref{eqn:es-ineqN}) to mixed states $\rho$, where
the formula (\ref{eqn:puretangle}) no longer valid is straightforward.
Given a pure state decomposition $\rho = \sum_i p_i |\psi_i\rangle \langle
\psi_i|$ we may define the average system-bath tangle by
\begin{equation}
    \bar{\tau}_{\mathcal{S}|\mathcal{B}}(\rho)  = \sum_i p_i \tau_{\mathcal{S}|\mathcal{B}} (|\psi_i\rangle) .
\end{equation}
The minimum $\bar{\tau}_{\mathcal{S}|\mathcal{B}}(\rho)$ over all pure-state
decompositions $\left\{ p_i, |\psi_i\rangle \right\}$ of $\rho$ can then
be used to quantify the quantum correlations between the system and the
bath.

The concavity of Eq.~(\ref{eqn:es-preineqN}) allows us to write
\begin{equation}
\frac{1}{N^2} \Bigg(1 +  \sqrt{1 -\tau^{\min}_{\mathcal{S}|\mathcal{B}}(\rho)} \Bigg)^2 \\
 \geq \sum_i p_i \tau_{B}(|\psi_i\rangle) .
\end{equation}
On the other hand the tangle is convex so we have
\(
    \sum_i p_i \tau_{B}(|\psi_i\rangle) \geq \tau_{B}(\rho) ,
\)
and thus obtain the following inequality
\begin{equation}
    \label{eqn:es-mixed}
\frac{1}{N^2}\left(1+\sqrt{1- \tau^{\min}_{\mathcal{S}|\mathcal{B}}(\rho)} \right)^2
    \geq \tau_{B}(\rho)
\end{equation}
which we can invert to obtain the entanglement-sharing inequality for
mixed states.


One simple model of decoherence where the inequality (\ref{eqn:es-ineqN}) is
immediately applicable is an exactly solvable model introduced by Zurek
\cite{Zurek82a} and recently used to investigate the structure of the
decoherence induced by spin environments \cite{ZCP03}. The system is
always in a pure state, so there are no classical correlations and a bound
on the system-bath entanglement is a bound on the decoherence.

The Hamiltonian of this model, after applying the complete symmetry
constraint, is written
\begin{equation}
H_{\mathcal{SB}} = \frac{1}{2}g \sum_{k=1}^{N}
\sigma_z^{(s)}\sigma^{(B_k)}_z .
\end{equation}
It is possible to analytically solve this model to give a good
illustration of how the decoherence of the central spin - as quantified by
the decay of the off-diagonal elements of the reduced density operator of
the system \cite{ZCP03} - is suppressed by the presence of entanglement
between the bath spins. Starting with a separable system-bath
($\mathcal{SB}$) state
\begin{equation}
|\Psi_{\mathcal{SB}}\rangle = \left(\chi |\downarrow\rangle_{\mathcal{S}} +\gamma |\uparrow\rangle_{\mathcal{S}} \right)\otimes |\mathcal{B}(0)\rangle,
\end{equation}
the state of $\mathcal{SB}$ at an arbitrary time $t$ is
\begin{equation}
|\Psi_{\mathcal{SB}}(t)\rangle = \chi |\downarrow\rangle_{\mathcal{S}} |\mathcal{B}_{\downarrow} (t)\rangle + \gamma |\uparrow\rangle_{\mathcal{S}} |\mathcal{B}_{\uparrow} (t)\rangle
\end{equation}
where
\begin{eqnarray}
|\mathcal{B}_{\downarrow}(t)\rangle = |\mathcal{B}_{\uparrow}(-t)\rangle = e^{igt\sum_{k=1}^{N}\sigma_z^{b_k}/2}|\mathcal{B}(0)\rangle.
\end{eqnarray}
The state of the system is then described by the reduced density operator,
\begin{eqnarray}
\rho_{\mathcal{S}} &=& |\chi|^2 |\downarrow\rangle_{\mathcal{S}}\langle \downarrow| +\chi\gamma^{*}r(t)|\downarrow\rangle_{\mathcal{S}}\langle \uparrow| \nonumber\\
&&+ \chi^{*}\gamma~r^{*}(t)|\uparrow\rangle_{\mathcal{S}}\langle\downarrow| + |\gamma|^2 |\uparrow\rangle_{\mathcal{S}}\langle \uparrow|
\end{eqnarray}
where the \emph{decoherence} factor \cite{ZCP03}, $r(t)=\langle\mathcal{B}_{\uparrow}(t)|\mathcal{B}_{\downarrow}(t)\rangle$ can be easily calculated. The absolute value of this factor is bounded by $0\leq |r(t)|^2 \leq 1$, corresponding to complete decoherence to a statistical mixture (0) and no loss of coherence (1), respectively. The $\mathcal{SB}$ tangle, $\tau_{\mathcal{S|B}}(t)$, can be written in terms of this factor by
\begin{equation}
\tau_{\mathcal{S|B}}(t) = 4|\chi|^2|\gamma|^2 \left(1-|r(t)|^2\right)
\end{equation}

We first consider  an initial bath state of the form
\begin{equation}
|\mathcal{B}(0)\rangle = \bigotimes_{k=1}^{N} \left(\alpha |\downarrow\rangle_{B_{k}} + \beta |\uparrow\rangle_{B_{k}}\right)
\end{equation}
which is completely separable, with each individual bath spin in an
identical state (preserving the symmetry). It is a relatively simple
exercise to calculate the decoherence factor,
\begin{equation}
|r(t)|^2 = \left[ |\alpha|^4 + |\beta|^4 + 2|\alpha|^2|\beta|^2 \cos(2gt)\right]^N.
\end{equation}
As argued in Zurek \emph{et. al.} \cite{ZCP03}, as $N\rightarrow\infty$,
the average value, $\langle|r(t)|^2\rangle \rightarrow 0$, implying
complete decoherence of the initial state. This is the average over time, since for large $N$, $|r(t)|^2$ is predominantly zero (over time) but will revive to one periodically. However as $N\rightarrow\infty$, these revival approach delta functions in time. With no intra-bath entanglement ($\tau_B=0$), there is no bound on
$\tau_{\mathcal{S|B}}$, resulting in maximal possible entanglement between
system and bath.  Unentangled baths of this form were the topic of Ref.
\cite{ZCP03}.

\begin{figure}[t]
\begin{center}
\scalebox{0.6}{\includegraphics{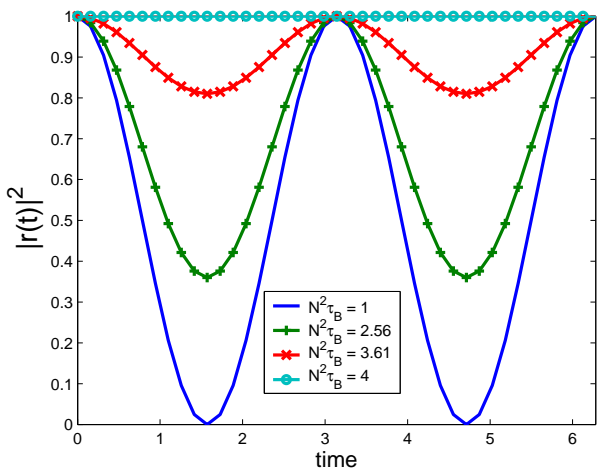}} \caption{(Color online) Plot of the
temporal evolution of the decoherence factor, $|r(t)|^2$ with an initial
entangled environmental state of the form Eq. (\ref{eqn:ent-env}) for different values of intra-bath tangle. We see that the entanglement in the bath acts to suppress the oscillation of $|r(t)|^2$, meaning the state of the system remains coherent.}
\label{fig:deco_fact}
\end{center}
\end{figure}

 We now consider an initial entangled environment state.
Following from the previous construction of the entanglement sharing
constraint, we choose an initial state of the form
\begin{eqnarray}
|\mathcal{B}(0)\rangle &=& \frac{a}{\sqrt{N}}\left(|\downarrow\downarrow\cdots\downarrow\downarrow\uparrow\rangle_{\mathcal{B}} +|\downarrow\downarrow\cdots\downarrow\uparrow\downarrow\rangle_{\mathcal{B}}\right. + \nonumber\\ &&\left. \cdots+|\uparrow\downarrow\ldots\downarrow\downarrow\downarrow\rangle_{\mathcal{B}}\right) + d|\downarrow\downarrow\cdots\downarrow\downarrow\rangle_{\mathcal{B}} \label{eqn:ent-env}
\end{eqnarray}
where $a^2 +d^2 =1$, such that the entanglement between any two bath spins
is $\tau_B = 4a^4$. Since the system-bath interaction does not flip spins, for such initial states the intra-bath entanglement is invariant over the evolution. In other words, the bath spins maintain their entanglement.  From this initial bath state, the decoherence factor is
\begin{equation}
|r(t)|^2 = |a|^4 + |d|^4 + 2|a|^2|d|^2 \cos(2gt),
\end{equation}
which, firstly, does not average to zero in the limit of large $N$ and, in fact, will not be zero at anytime for
given values of $a$ and $d$ (see Figure \ref{fig:deco_fact}). This can be interpreted as a suppression of
decoherence, since at no time will the system ever be a complete
statistical mixture of states.

The inequality only places a nontrivial upper bound on the system-bath entanglement when $\tau_B \geq 1/N^2$. For the states
considered here, this corresponds to the parameter range $1/\sqrt{2} \leq
a\leq 1$, to which we will now restrict ourselves. The system-bath
tangle is given by
\begin{equation}
\tau_{\mathcal{S|E}} = 2|a|^2 (1-|a|^2)(1-\cos(2gt)).
\end{equation}
From the intra-bath tangle, $\tau_B = 4a^4$, the entanglement sharing
inequality (\ref{eqn:es-ineqN}) gives an upper bound on the
system-bath tangle of
\begin{equation}
\tau^{\mathrm{max}}_{\mathcal{S|E}} = 4|a|^2\left(1-|a|^2\right)
\end{equation}
and it is simple to show that, $\tau_{\mathcal{S|E}}\leq
\tau^{\mathrm{max}}_{\mathcal{S|E}}$. In turn, this constrains the lower
bound on the decoherence factor. This simple example demonstrates that
entanglement in the environment can constrain entanglement between the
system and environment, and hence limit the effect of decoherence. Of course, in this example we have not considered any intrinsic central spin or bath dynamics.

It is also possible to calculate the intra-bath entanglement for the
Tessieri-Wilkie model \cite{Tessieri02a}, where the initial state of the bath is a thermal
state and thus the overall state at time $t$ is mixed.
\begin{figure}[t]
\begin{center}
\scalebox{0.5}{\includegraphics{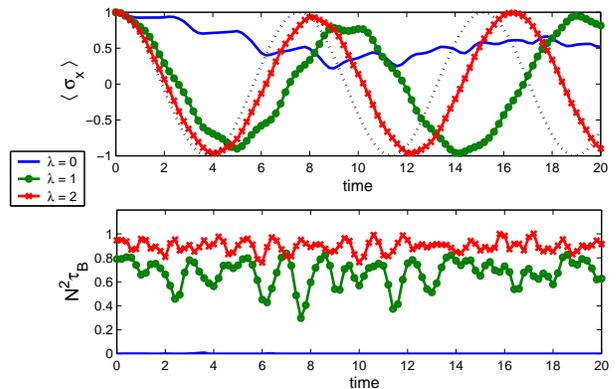}} \caption{(Color online) Rabi oscillations and the intra-bath entanglement, quantified by the tangle between any two bath spins,  for three different intra-bath coupling strengths for the Tessieri and Wilkie model, with $N=10$ bath spins. The dotted line in the $\langle \hat{\sigma}_x\rangle$ plot is the case of no system-bath interaction. As the intra-bath coupling increases, so does the intra-bath entanglement, and the Rabi oscillations approach the limit of no system-bath interaction(dotted line).}
\label{fig:TW_Rabi}
\end{center}
\end{figure}
In the Tessieri-Wilkie model, the system is described by
\begin{equation}
H_{\mathcal{S}} = \frac{\omega_0}{2}\sigma_z^{(0)} + \beta \sigma_x^{(0)},
\end{equation}
the bath,
\begin{equation}
H_{\mathcal{B}} = \sum_{i=1}^N \frac{\omega_i}{2}\sigma_z^{(i)} + \beta\sum_{i=1}^N \sigma_x^{(i)} + \lambda \sum_{i=1}^{N-1} \sum_{j=1}^N \sigma_x^{(i)}\sigma_x^{(j)},
\end{equation}
and the interaction,
\begin{equation}
H_{\mathcal{SB}} = \lambda_0\sum_{i=1}^N \sigma_x^{(i)}\sigma_x^{(0)}.
\end{equation}
Following Ref. \cite{Tessieri02a}, $\beta = 0.01$, $\lambda_0 = 1$ and $\omega_0 = 0.8288$, however we set $\omega_i = 1$ such that all baths spins are identical. The bath starts in the thermal state, $\rho_{\mathcal{B}}(0) = \exp{(-H_{\mathcal{B}}/kT)} / \left(\textrm{Tr}[\exp{(-H_{\mathcal{B}}/kT)}]\right)$, such that varying the intra-bath coupling strength $\lambda$, varies the initial entanglement between the bath spins. To see the effects of decoherence, the central spin is initialized in the state $|\psi_{\mathcal{S}}(0)\rangle = \left(|\uparrow\rangle +|\downarrow\rangle\right)/\sqrt{2}$. In the absence of the bath, the central spin will simply precess, exhibiting Rabi oscillations. Interactions with the bath that decohere the spin will prevent such coherent oscillations.

Figure \ref{fig:TW_Rabi} shows how an entangled bath can suppress the decohering effects of the bath, allowing coherent oscillations of the central spin. Since the bath begins in thermal equilibrium, it's state does not vary significantly over its evolution (especially if $N$ is large). Hence, if the initial state is entangled, this degree of entanglement is maintained throughout the evolution.

Since the initial state is mixed, classical correlations between system and bath will be a cause of decoherence. However it is likely that the result of
Koashi \cite{Koashi03a} may be extended to the central-spin model, thus
showing that suppression of decoherence is a generic feature when
spin-bath environments maintain a high degree of internal entanglement.

In order to gain insight into how intra-bath entanglement can reduce
decoherence we have
considered two simple models in which all bath spins interact equally
with one another.
This represents a model for which mean field approximation for the
interaction between spins is exact. More physical models will
involve short range interactions, yet we conjecture that they will exhibit essentially the same phenomena.

Recent studies of a central spin/qubit interacting with a reservoir of (identical) qubits has considered the process of \emph{homogenization} \cite{ZSB+02}, of which thermalization is a special case \cite{SZS+02}. The system qubit is initially in some state $\rho$, with the all bath spins each in the identical state, $\xi$. The aim of the process is to output all qubits in some arbitrarily small neighborhood of $\xi$. Thermalization is the case were $\xi$ corresponds to the thermal state. This thermalization process is equivalent to the decoherence of the system qubit to a thermal state.

In this discrete time process, the system qubit interacts with a only single bath qubit at each time step, and never the same qubit twice. It is shown that the partial swap operation uniquely determines a universal quantum homogenizer \cite{ZSB+02}. While there is no explicit interaction between bath qubits, their mutual interaction with the system qubit generates entanglement not only between the system and reservoir, but also intra-bath entanglement. This entanglement is studied in \cite{ZSB+02} and the results agree with the entanglement sharing arguments we have made here. Specifically, in the example considered, the entanglement between system and bath decreases in the long term, as more bath qubits become entangled with each other. Interestingly, it is shown that all entanglements are pairwise, with no multi-party entanglement present \cite{ZSB03}. It would be interesting to extend the work in these articles by considering thermalization in the presence of a self-interacting bath. Of course, different methods would have to be employed, since the state of the bath qubits would change after each interaction.

Decoherence is the major stumbling block on the road to quantum computing. Here we have introduced a novel way of constraining the decoherence effects from a spin-bath environment. Such environmental models are of particular importance for predicting decoherence effects in solid-state qubits in the low temperature regime \cite{DS01,Sta03}.

We have used two simplified models as examples of how entanglement in the environmental bath may suppress decoherence. While we have only discussed spin-baths, one could also envision similar effects for oscillator baths, where entangled spins may be replaced by multi-mode squeezed states. As well we have focussed upon two-party entanglement in the bath. The effects of $m$-party entangled states may be quite different.

The types of entangled states of the bath that may be created and maintained will depend explicitly upon the physical system in question. To discover if entanglement-sharing can suppress decoherence in realistic situations requires calculations for specific quantum computer architectures. Only then will it be apparent if this unique property of entanglement can be used to our advantage in overcoming decoherence. \\

\acknowledgments

We thank Michael Nielsen for helpful discussions on entanglement-sharing inequalities. APH thanks Philip Stamp for enjoyable and enlightening discussions about the `real world' of spin-baths. This work was supported by the Australian Research Council as part of the Centre of Excellence for Quantum Computer Technology.

\bibliography{ent_spin_bath}

\begin{thebibliography}{20}
\expandafter\ifx\csname natexlab\endcsname\relax\def\natexlab#1{#1}\fi
\expandafter\ifx\csname bibnamefont\endcsname\relax
  \def\bibnamefont#1{#1}\fi
\expandafter\ifx\csname bibfnamefont\endcsname\relax
  \def\bibfnamefont#1{#1}\fi
\expandafter\ifx\csname citenamefont\endcsname\relax
  \def\citenamefont#1{#1}\fi
\expandafter\ifx\csname url\endcsname\relax
  \def\url#1{\texttt{#1}}\fi
\expandafter\ifx\csname urlprefix\endcsname\relax\def\urlprefix{URL }\fi
\providecommand{\bibinfo}[2]{#2}
\providecommand{\eprint}[2][]{\url{#2}}

\bibitem[{\citenamefont{Bennett and DiVincenzo}(2000)}]{Bennett00a}
\bibinfo{author}{\bibfnamefont{C.}~\bibnamefont{Bennett}} \bibnamefont{and}
  \bibinfo{author}{\bibfnamefont{D.}~\bibnamefont{DiVincenzo}},
  \bibinfo{journal}{Nature} \textbf{\bibinfo{volume}{404}},
  \bibinfo{pages}{247} (\bibinfo{year}{2000}).

\bibitem[{\citenamefont{Zurek}(1991)}]{Zurek91a}
\bibinfo{author}{\bibfnamefont{W.}~\bibnamefont{Zurek}},
  \bibinfo{journal}{Phys. Today} \textbf{\bibinfo{volume}{44}},
  \bibinfo{pages}{36} (\bibinfo{year}{1991}).

\bibitem[{\citenamefont{Zurek}(1982)}]{Zurek82a}
\bibinfo{author}{\bibfnamefont{W.}~\bibnamefont{Zurek}},
  \bibinfo{journal}{Phys. Rev. D} \textbf{\bibinfo{volume}{26}},
  \bibinfo{pages}{1862} (\bibinfo{year}{1982}).

\bibitem[{\citenamefont{Weiss}(1999)}]{Weiss99a}
\bibinfo{author}{\bibfnamefont{U.}~\bibnamefont{Weiss}},
  \emph{\bibinfo{title}{Quantum dissipative systems}}
  (\bibinfo{publisher}{World Scientific, Singapore}, \bibinfo{year}{1999}),
  \bibinfo{edition}{2nd} ed.

\bibitem[{\citenamefont{Prokof'ev and Stamp}(2000)}]{Prokofev00a}
\bibinfo{author}{\bibfnamefont{N.}~\bibnamefont{Prokof'ev}} \bibnamefont{and}
  \bibinfo{author}{\bibfnamefont{P.}~\bibnamefont{Stamp}},
  \bibinfo{journal}{Rep. Prog. Phys} \textbf{\bibinfo{volume}{63}},
  \bibinfo{pages}{669} (\bibinfo{year}{2000}).

\bibitem[{\citenamefont{Tessieri and Wilkie}(2002)}]{Tessieri02a}
\bibinfo{author}{\bibfnamefont{L.}~\bibnamefont{Tessieri}} \bibnamefont{and}
  \bibinfo{author}{\bibfnamefont{J.}~\bibnamefont{Wilkie}},
  \bibinfo{journal}{Journal of Physics A} \textbf{\bibinfo{volume}{36}},
  \bibinfo{pages}{12305} (\bibinfo{year}{2002}).

\bibitem[{\citenamefont{Paganelli et~al.}(2002)\citenamefont{Paganelli,
  de~Pasquale, and Giampaolo}}]{Paganelli02a}
\bibinfo{author}{\bibfnamefont{S.}~\bibnamefont{Paganelli}},
  \bibinfo{author}{\bibfnamefont{F.}~\bibnamefont{de~Pasquale}},
  \bibnamefont{and} \bibinfo{author}{\bibfnamefont{S.~M.}
  \bibnamefont{Giampaolo}}, \bibinfo{journal}{Phys. Rev. A}
  \textbf{\bibinfo{volume}{66}}, \bibinfo{pages}{052317}
  (\bibinfo{year}{2002}).

\bibitem[{\citenamefont{Lucamarini et~al.}(2004)\citenamefont{Lucamarini,
  Paganelli, and Mancini}}]{Lucamarini04a}
\bibinfo{author}{\bibfnamefont{M.}~\bibnamefont{Lucamarini}},
  \bibinfo{author}{\bibfnamefont{S.}~\bibnamefont{Paganelli}},
  \bibnamefont{and} \bibinfo{author}{\bibfnamefont{S.}~\bibnamefont{Mancini}},
  \bibinfo{journal}{Phys. Rev. A} \textbf{\bibinfo{volume}{69}},
  \bibinfo{pages}{062308} (\bibinfo{year}{2004}),
  \bibinfo{note}{{arXiv}:quant-ph/0402073}.

\bibitem[{\citenamefont{Coffman et~al.}(2000)\citenamefont{Coffman, Kundu, and
  Wootters}}]{Coffman00a}
\bibinfo{author}{\bibfnamefont{V.}~\bibnamefont{Coffman}},
  \bibinfo{author}{\bibfnamefont{J.}~\bibnamefont{Kundu}}, \bibnamefont{and}
  \bibinfo{author}{\bibfnamefont{W.~K.} \bibnamefont{Wootters}},
  \bibinfo{journal}{Phys. Rev. A} \textbf{\bibinfo{volume}{61}},
  \bibinfo{pages}{052306} (\bibinfo{year}{2000}),
  \bibinfo{note}{{arXiv}:quant-ph/9907047}.

\bibitem[{\citenamefont{Koashi and Winter}(2003)}]{Koashi03a}
\bibinfo{author}{\bibfnamefont{M.}~\bibnamefont{Koashi}} \bibnamefont{and}
  \bibinfo{author}{\bibfnamefont{A.}~\bibnamefont{Winter}},
  \bibinfo{journal}{Phys. Rev. A} \textbf{\bibinfo{volume}{69}},
  \bibinfo{pages}{022309} (\bibinfo{year}{2003}),
  \bibinfo{note}{{arXiv}:quant-ph/0310037}.

\bibitem[{\citenamefont{Wootters}(1998)}]{Wootters98a}
\bibinfo{author}{\bibfnamefont{W.~K.} \bibnamefont{Wootters}},
  \bibinfo{journal}{Phys. Rev. Lett.} \textbf{\bibinfo{volume}{80}},
  \bibinfo{pages}{2245} (\bibinfo{year}{1998}).

\bibitem[{\citenamefont{D\"ur et~al.}(2000)\citenamefont{D\"ur, Vidal, and
  Cirac}}]{Dur00a}
\bibinfo{author}{\bibfnamefont{W.}~\bibnamefont{D\"ur}},
  \bibinfo{author}{\bibfnamefont{G.}~\bibnamefont{Vidal}}, \bibnamefont{and}
  \bibinfo{author}{\bibfnamefont{J.~I.} \bibnamefont{Cirac}},
  \bibinfo{journal}{Phys. Rev. A} \textbf{\bibinfo{volume}{62}},
  \bibinfo{pages}{062314} (\bibinfo{year}{2000}),
  \bibinfo{note}{{arXiv}:quant-ph/0005115}.

\bibitem[{\citenamefont{Carteret et~al.}(2000)\citenamefont{Carteret, Higuchi,
  and Sudbery}}]{Carteret00a}
\bibinfo{author}{\bibfnamefont{H.~A.} \bibnamefont{Carteret}},
  \bibinfo{author}{\bibfnamefont{A.}~\bibnamefont{Higuchi}}, \bibnamefont{and}
  \bibinfo{author}{\bibfnamefont{A.}~\bibnamefont{Sudbery}},
  \bibinfo{journal}{J. Math. Phys.} \textbf{\bibinfo{volume}{41}},
  \bibinfo{pages}{7932} (\bibinfo{year}{2000}),
  \bibinfo{note}{{arXiv}:quant-ph/0006125}.

\bibitem[{\citenamefont{Koashi et~al.}(2000)\citenamefont{Koashi, Buzek, and
  Imoto}}]{Koashi00a}
\bibinfo{author}{\bibfnamefont{M.}~\bibnamefont{Koashi}},
  \bibinfo{author}{\bibfnamefont{V.}~\bibnamefont{Buzek}}, \bibnamefont{and}
  \bibinfo{author}{\bibfnamefont{N.}~\bibnamefont{Imoto}},
  \bibinfo{journal}{Phys. Rev. A} \textbf{\bibinfo{volume}{62}},
  \bibinfo{pages}{050302(R)} (\bibinfo{year}{2000}),
  \bibinfo{note}{{arXiv}:quant-ph/0007086}.

\bibitem[{\citenamefont{Zurek et~al.}(2003)\citenamefont{Zurek, Cucchietti, and
  Paz}}]{ZCP03}
\bibinfo{author}{\bibfnamefont{W.~H.} \bibnamefont{Zurek}},
  \bibinfo{author}{\bibfnamefont{F.~M.} \bibnamefont{Cucchietti}},
  \bibnamefont{and} \bibinfo{author}{\bibfnamefont{J.~P.} \bibnamefont{Paz}}
  (\bibinfo{year}{2003}), \bibinfo{note}{{arXiv}:quant-ph/0312207}.

\bibitem[{\citenamefont{Ziman et~al.}(2002)\citenamefont{Ziman, {P. \u
  Stlmachovi\u c}, {V. Bu\u zek}, Hillary, Scarani, and Gisin}}]{ZSB+02}
\bibinfo{author}{\bibfnamefont{M.}~\bibnamefont{Ziman}},
  \bibinfo{author}{\bibnamefont{{P. \u Stlmachovi\u c}}},
  \bibinfo{author}{\bibnamefont{{V. Bu\u zek}}},
  \bibinfo{author}{\bibfnamefont{M.}~\bibnamefont{Hillary}},
  \bibinfo{author}{\bibfnamefont{V.}~\bibnamefont{Scarani}}, \bibnamefont{and}
  \bibinfo{author}{\bibfnamefont{N.}~\bibnamefont{Gisin}},
  \bibinfo{journal}{Phys. Rev. A} \textbf{\bibinfo{volume}{65}},
  \bibinfo{pages}{042105} (\bibinfo{year}{2002}).

\bibitem[{\citenamefont{Scarani et~al.}(2002)\citenamefont{Scarani, Ziman, {P.
  \u Stelmachovi\u c}, Gisin, and {V. Bu\u zek}}}]{SZS+02}
\bibinfo{author}{\bibfnamefont{V.}~\bibnamefont{Scarani}},
  \bibinfo{author}{\bibfnamefont{M.}~\bibnamefont{Ziman}},
  \bibinfo{author}{\bibnamefont{{P. \u Stelmachovi\u c}}},
  \bibinfo{author}{\bibfnamefont{N.}~\bibnamefont{Gisin}}, \bibnamefont{and}
  \bibinfo{author}{\bibnamefont{{V. Bu\u zek}}}, \bibinfo{journal}{Phys. Rev.
  Lett.} \textbf{\bibinfo{volume}{88}}, \bibinfo{pages}{097905}
  (\bibinfo{year}{2002}).

\bibitem[{\citenamefont{Ziman et~al.}(2003)\citenamefont{Ziman, {P. \u
  Stelmachovi\u c}, and {V. Bu\u zek}}}]{ZSB03}
\bibinfo{author}{\bibfnamefont{M.}~\bibnamefont{Ziman}},
  \bibinfo{author}{\bibnamefont{{P. \u Stelmachovi\u c}}}, \bibnamefont{and}
  \bibinfo{author}{\bibnamefont{{V. Bu\u zek}}}, \bibinfo{journal}{J. Opt. B:
  Quantum Semiclass. Opt.} \textbf{\bibinfo{volume}{5}}, \bibinfo{pages}{S439}
  (\bibinfo{year}{2003}).

\bibitem[{\citenamefont{Dub\'e and Stamp}(2001)}]{DS01}
\bibinfo{author}{\bibfnamefont{M.}~\bibnamefont{Dub\'e}} \bibnamefont{and}
  \bibinfo{author}{\bibfnamefont{P.~C.~E.} \bibnamefont{Stamp}},
  \bibinfo{journal}{Chem. Phys.} \textbf{\bibinfo{volume}{268}},
  \bibinfo{pages}{257} (\bibinfo{year}{2001}).

\bibitem[{\citenamefont{Stamp}(2003)}]{Sta03}
\bibinfo{author}{\bibfnamefont{P.~C.~E.} \bibnamefont{Stamp}},
  \bibinfo{journal}{J. Quantum Computers and Computing}
  \textbf{\bibinfo{volume}{4}}, \bibinfo{pages}{20} (\bibinfo{year}{2003}).

\end{thebibliography}

\end{document}